\documentclass{teormf-j}
\usepackage{graphicx} 
\usepackage{tabularx}

\theoremstyle{definition}

\theoremstyle{remark}

\numberwithin{equation}{section}

\begin{document}
\title{Biefeld{}-Brown effect and space curvature of electromagnetic field
}


\author{A.Maknickas 
}
\email{alm@sc.vgtu.lt}



\date{\today}

\begin{abstract}
With applying of new proposed electromagnetic gravity Lagrangian together with Einstein-Hilbert equation not zero space curvature was derived. The curvature gives ``a priory'' postulate of equivalence of mass and electromagnetic field gravity properties. The non zero trace of energy-stress tensor of electrical field  changes space curvature of gravity mass, which yields to prediction of dependence of capacitor gravity mass from capacitor capacitance and voltage values, observed in Biefeld{}-Brown effect. The other, not observed prediction could be applied to coil gravity mass dependence from coil inductance and current values. New physical constant, electromagnetic field gravity constant ${\alpha_g}$, was introduced to conform with theoretical and experimental data.
\end{abstract}
\maketitle
\section{Introduction}
\label{intro}
Biefeld{}-Brown \cite{Ref1}-\cite{Ref6} effect has been known since 1928. \ Christensen and M{\o}ller \cite{Ref7} built a Biefeld{}-Brown electrode setup and published measurements of the obtained thrust in ambient air. They also compared their results with theoretical predictions of electric wind effects. The agreement was very good and tended to explain the Biefeld{--}Brown effect as a purely electric wind phenomenon. \ The other try to verify the adequacy of Biefeld{}-Brown effect with electric wind phenomena was made by Tajmar \cite{Ref8}. The author concluded in this article, that electric wind phenomena explain Biefeld{}-Brown effect. \ The results \cite{Ref8} suggest that corona wind effects was misinterpreted as a possible connection between gravitation and electromagnetism. Brown observed in \cite{Ref9} that this effect remained even if the ambient medium were a vacuum (up to 10\textsuperscript{$-$}\textsuperscript{6} Torr). \ Talley \cite{Ref10}, \cite{Ref11} studied Brown's electrode \ configurations in vacuum chambers up to 10\textsuperscript{$-$}\textsuperscript{6} Torr in great detail. He
found no thrust in the case of a static dc potential applied to the
electrodes. However, he noticed an anomalous force during electrical
breakdowns when the current was flowing. This force, the result of currents
in divergent electric fields (due to the asymmetrical electrode
configuration) finds further support in five{}-dimensional theories
coupling the gravitational and electromagnetic field. Williams \cite{Ref12}
integrated a mass dependent fifth dimension into the relativistic
Maxwell theory and predicted the coupling between both fields. However, ambiguity of explanation of Biefeld{}-Brown effect indicates that new
theoretical and experimental researches are needed.
\section{Einstein's field equations}
\label{sec:a1}
Suppose that the full action of the theory is given by the Einstein-Hilbert \cite{Ref122} term plus a term ${\mathcal{L}_\mathrm{M}}$ describing any matter fields appearing in the theory
\begin{align}
S = \int \left[ \frac{1}{2\kappa} \, R + \mathcal{L}_\mathrm{M} \right] \sqrt{-g} \, \mathrm{d}^4 x
\end{align}

The action principle then tells us that the variation of this action with respect to the inverse metric is zero, yielding 
\begin{align}
0 & = \delta S \\
  & = \int 
        \left[ 
           \frac{1}{2\kappa} \left( \frac{\delta R}{\delta g^{\mu\nu}} +
             \frac{R}{\sqrt{-g}} \frac{\delta \sqrt{-g}}{\delta g^{\mu\nu} } 
            \right) + 
           \frac{1}{\sqrt{-g}} \frac{\delta (\sqrt{-g} \mathcal{L}_\mathrm{M})}{\delta g^{\mu\nu}} 
        \right] \delta g^{\mu\nu} \sqrt{-g}\, \mathrm{d}^4x.
\end{align}
Since this equation should hold for any variation ${\delta g^{\mu\nu}}$, it implies that
\begin{align}
\frac{\delta R}{\delta g^{\mu\nu}} + \frac{R}{\sqrt{-g}} \frac{\delta \sqrt{-g}}{\delta g^{\mu\nu}} 
= - 2 \kappa \frac{1}{\sqrt{-g}}\frac{\delta (\sqrt{-g} \mathcal{L}_\mathrm{M})}{\delta g^{\mu\nu}},
\end{align}
this equation of motion for the metric field could be found. The calculation of the left hand side of the equation for the variations of the Ricci scalar R and the determinant of the metric could be found in Carroll \cite{Ref0}. After having of all the necessary variations at our disposal, we can insert them into the equation of motion for the metric field to obtain
\begin{align}
R_{\mu\nu} - \frac{1}{2} g_{\mu\nu} R = \frac{8 \pi G}{c^4}  T_{\mu\nu}, \label{eq:motion}
\end{align}
which is Einstein's field equation ~\cite{Ref13}, \cite{Ref14} and constant
\begin{align}
\kappa = \frac{8 \pi G}{c^4}
\end{align}
has been chosen so that the non-relativistic limit yields the usual form of Newton's gravity law, where ${G}$ is the gravitational constant and ${c}$ is speed of light in vacuum. The right hand side of this equation is (by definition) proportional to the energy-stress tensor
\begin{equation}
\label{eq:enstressten}
\begin{aligned}
T_{\mu\nu} = -2 \frac{1}{\sqrt{-g}}\frac{\delta (\sqrt{-g} \mathcal{L}_\mathrm{M})}{\delta g^{\mu\nu}} 
= -2 \frac{\delta \mathcal{L}_\mathrm{M}}{\delta g^{\mu\nu}} + g_{\mu\nu} \mathcal{L}_\mathrm{M}.
\end{aligned}
\end{equation}
The Lagrangian of mater must be chosen so, that it satisfy conservation lows.

\section{Lagrangian of electromagnetic field}
The electromagnetic tensor ${F^{\mu\nu}}$ in Cartesian coordinates is commonly written as a matrix: 
\begin{align}
F^{\mu\nu} = \begin{bmatrix}
0 & -E_x/c & -E_y/c & -E_z/c \\
E_x/c & 0 & -B_z & B_y \\
E_y/c & B_z & 0 & -B_x \\
E_z/c & -B_y & B_x & 0
\end{bmatrix},
\end{align} 
or
\begin{align}
F_{\mu\nu} = \begin{bmatrix}
0 & E_x/c & E_y/c & E_z/c \\
-E_x/c & 0 & -B_z & B_y \\
-E_y/c & B_z & 0 & -B_x \\
-E_z/c & -B_y & B_x & 0
\end{bmatrix},
\end{align} 
where ${E}$ is the electric field, ${B}$ the magnetic field, and ${c}$ the speed of light. The signs in the tensor above depend on the convention used for the metric tensor. The convention used here is ${+ - - -}$, corresponding to the metric tensor:
\begin{align}
\begin{pmatrix} 1 & 0 & 0 & 0 \\ 0 & -1 & 0 & 0 \\ 0 & 0 & -1 & 0 \\ 0 & 0 & 0 & -1 \end{pmatrix}.
\end{align}

From the matrix form of the field tensor, it becomes clear that the electromagnetic tensor satisfies the following antisymmetry properties 
 ${F^{ab} \, = - F^{ba}}$ (hence the name bi vector) of six independent components.

If one forms an inner product of the field strength tensor Lorentz invariant is formed:
\begin{align}
F_{ab} F^{ab} = \ 2 \left( B^2 - \frac{E^2}{c^2} \right) = \mathrm{invariant}.
\end{align}
The Lagrangian of electromagnetic field in our model could be
\begin{align}
\mathcal{L}_{em} = -\frac{ \alpha_g c^2 }{4\mu_0} F_{ab} F^{ab}. 
\end{align}
Lagrangian ${\mathcal{L}_{em}}$ differs from classic electromagnetic field Lagrangian just with constant ${\alpha_g c^2}$, where ${\alpha_g}$ is electromagnetic field gravity constant with dimension $\frac{s^2}{m^{2}}$, which could be calibrated on experiment data.

According to ~\cite{Ref14_1}, it is safer to rewrite inner product of field strength tensor as ${F_{ab}F_{cd}g^{ac}g^{bd}}$. This gives for the first term of sum of energy-stress tensor eq. ~\eqref{eq:enstressten}
\begin{align}
\label{eq:eq36}
2 \frac{\delta \mathcal{L}_{em}}{\delta g^{\mu\nu}} = - 4 F_{l \nu} F_{\mu}^l ,
\end{align}
A different rezult could be obtaned, if inner product of field strength tensor is leaving as ${F_{ab}F^{ab}}$. In this way result is:
\begin{align}
\label{eq:eq37}
2 \frac{\delta \mathcal{L}_{em}}{\delta g^{\mu\nu}} = 0.
\end{align}
Let's deside, that ~\eqref{eq:eq37} is true\footnote{discussion in A\textsc{ppendix}}, so the calculations of energy-stress tensor term of electromagnetic field give result:
\begin{equation}
T_{\mu\nu}^{\left(em\right)} = -2 \frac{\delta \mathcal{L}_{em}}{\delta g^{\mu\nu}} + g_{\mu\nu} \mathcal{L}_{em} =  -g_{\mu\nu} \frac{\alpha_g c^2}{2\mu_0} \left( B^2 - \frac{E^2}{c^2} \right),
\end{equation}
which is symmetric and satisfy rotation conservation low.
 
\section{Einstein's field equation for gravity mass in electromagnetic field}
\label{sec:1}
According to ~\eqref{eq:enstressten} energy-stress tensor of gravity mass in electromagnetic field could be found with adding Lagrangian of gravitational and electromagnetic parts of Lagrangian   
\begin{subequations}
\begin{align}
\mathcal{L}_\mathrm{M} &= \mathcal{L}_{\rho} + \mathcal{L}_{em} = -\rho \mathrm{c}^2 - \frac{\alpha_g c^2}{4\mu_0} F_{\mu\nu} F^{\mu\nu},\\
T_{\mu\nu}&= g_{\mu\nu} \mathcal{L}_\mathrm{M}, \\ 
&= -g_{\mu\nu} \rho \mathrm{c}^2 -g_{\mu\nu} \frac{\alpha_g c^2}{2\mu_0} \left( B^2 - \frac{E^2}{c^2} \right). \label{eq:Eq26b}
\end{align} 
\end{subequations}
After inserting of energy-stress tensor ~\eqref{eq:Eq26b} into ~\eqref{eq:motion} Einstein-Hilbert field equation looks like this
\begin{align}
R_{\mu\nu} - \frac{1}{2} g_{\mu\nu} R =-g_{\mu\nu} \frac{8 \pi G}{c^4}  \left( \rho \mathrm{c}^2 + \frac{ \alpha_g c^2 }{2\mu_0} \left( B^2 - \frac{E^2}{c^2} \right) \right). \label{eq:metricef}
\end{align}
Taking the trace of ~\eqref{eq:metricef} (contracting with ${g^{\mu \nu}}$) and using the fact that ${g^{\mu \nu} g_{\mu \nu} = 4}$, we get for space curvature:
\begin{align}
R = \frac{32 \pi G}{c^4}  \left( \rho {c}^2 + \frac{ \alpha_g c^2 }{2\mu_0} \left( B^2 - \frac{E^2}{c^2} \right) \right), \label{eq:curvature}
\end{align}
yielding the equivalent form of ~\eqref{eq:metricef}
\begin{align}
R_{\mu \nu} = g_{\mu \nu} \frac{8 \pi G}{c^4}  \left( \rho {c}^2 + \frac{ \alpha_g c^2 }{2\mu_0} \left( B^2 - \frac{E^2}{c^2} \right) \right),
\end{align}
Space curvature ~\eqref{eq:curvature} of spheric gravity mass with radius ${r}$ in electromagnetic field could be rewritten as
\begin{align}
\label{eq:Eq16}
R &= \frac{24 G}{{c}^2 r^3} \left( {\rho} - {\rho_{eg}} \right), \\ 
\label{eq:Eq17}
M_{eg} &= \rho_{eg} V = \frac{ \alpha_g V}{2} \left( \varepsilon_0 E^2 - \frac{B^2}{\mu_0} \right), \\  
V &= \frac{4 \pi r^3}{3},
\end{align}
where ${M_{eg}}$ is electromagnetic mass and ${V}$ is volume of electromagnetic field and is equal to volume of devices, which is inside this electromagnetic field.

From eq.~\eqref{eq:Eq16} assumption could be made, that summary curvature of the space
generated by gravity mass should decrease in electric field and increase in magnetic field. If the mass of device ${M}$ equals to absolute value of electromagnetic mass ${\left|M_{eg}\right|}$, zero curvature of such device could be obtained, so gravity mass of this device stop interacting with the other gravity mass. Let's prove it analytically.

The solution of eq.~\eqref{eq:Eq16} in spheric coordinates (see for example \cite{Ref19}) is: 
\begin{align}
g_{\mu \nu} = \begin{pmatrix} 1-\frac{\Lambda r^2}{3} & 0 & 0 & 0 \\ 0 & -\frac{1}{1-\frac{\Lambda r^2}{3}} & 0 & 0 \\ 0 & 0 & -r^2 & 0 \\ 0 & 0 & 0 & -r^2 \sin{\theta} \end{pmatrix}, 
\end{align}
where ${\Lambda}$ equals to 
\begin{equation}
\Lambda = \frac{8 \pi G}{c^4}  \left( \rho {c}^2 + \frac{ \alpha_g c^2 }{2\mu_0} \left( B^2 - \frac{E^2}{c^2} \right) \right).
\end{equation}
Thus, the gravitational potential of a point mass is
\begin{align}
\Phi &= - \frac{\Lambda r^2 c^2}{6}\\ 
&= - \frac{4 \pi G r^2}{3}  \left( \rho + \frac{ \alpha_g }{2\mu_0} \left( B^2 - \frac{E^2}{c^2} \right) \right) \\
&=-\frac{G \left(M - M_{eg}\right)}{r}. \label{eq:potential}.
\end{align}
Eq.~\eqref{eq:potential} prove proposition, if the mass of device ${M}$ equals to absolute value of ${\left|M_{eg}\right|}$, we have zero gravity potential of device with gravity mass ${M}$ and such device does not interact with gravity field of the other gravity mass.
\section{Biefeld{}-Brown effect}
\label{sec:2}
Replacing of magnetic field ${B}$ with ${B=0}$ and replacing of electric field density with capacitor energy density multiplied by volume of electric field of capacitor in eq. ~\eqref{eq:Eq17} gives equation for electro gravity mass of capacitor
\begin{equation}
\label{eq:Eq20}
M_{eg}=\frac{C U^2}{2}{\alpha_g}.
\end{equation}
The simple capacitor of two parallel conductive plates describes as 
\begin{equation}
C = \frac{\varepsilon \varepsilon_{0} S}{d},
\end{equation}
so eq.~\eqref{eq:Eq20} could be rewritten as: 
\begin{equation}
\label{eq:Eq21}
M_{{{{eg}}}}=\frac{{{\varepsilon \varepsilon}}_{{0}}{{SU}}^{{2}}}{2 d} \alpha_g,
\end{equation}
where ${\varepsilon}$ is relative permittivity of material,
${\varepsilon_0}$ is absolute permittivity of vacuum, S
is the area of the capacitor and d is separation of the planes of the
capacitor. Eq.~\eqref{eq:Eq21} explains all experimental data observed Biefeld{}-Brown
effect. The effect depends on the separation of
the plates of the capacitor, the closer the plates, the greater the
effect. The effect depends on the dielectric strength of the
material between the electrodes, the higher the strength, the greater
the effect. The effect depends on the area of the conductor, the
greater the area, the greater the effect. The effect depends on the
voltage difference between the plates, the greater the voltage, the
greater effect. The effect depends on the volume of the dielectric
material, the greater the volume, the greater part of electromagnetic
energy concentrates in dielectric material, the
greater the effect.

Electro-gravity constant ${\alpha_g}$ could be found from White et. al \cite{Ref31} experiments. Authors measured in their experiments  propulsion of asymmetric radio frequency (RF)  electric field resonator force in vacuum
\begin{equation}
\label{eq:Eq23}
\Delta F   = \Delta m g = \alpha_g V \frac{\epsilon_0 E^2}{2} g = \alpha_g P \Delta  t g,
\end{equation}
where ${\Delta F}$ is lifting force,  $g$ is gravity acceleration constant, $V$ resonator volume, $E$ electric field strength,  $P$ applied power in resonator, $t$ power impulse duration and constant ${\alpha_g}$ is electro-gravity constant. The inserting of numeric values (Tab. 1)  into \eqref{eq:Eq23} and calculating least squire line (Fig. 1) slope divided be $g=9.80065$ results for electro-gravity constant  in value ${\alpha_g = 2.60\pm 0.47E-9}$ [$\frac{s^2}{m^{2}}$] .
\begin{table}[h]
\caption{RF resonator trust force  experiment data \cite{Ref31}}
\begin{tabular}{|c|c|c|c|}
\hline 
Power, W & Duration, s & Energgy, J & Force, $\mu$N\tabularnewline
\hline 
\hline 
41.4  & 23.0  & 952.20  & 30 \tabularnewline
\hline 
60.2  & 22.2  & 1336.44  & 43 \tabularnewline
\hline 
40.8  & 39.8  & 1623.84  & 30 \tabularnewline
\hline 
55.0  & 33.0  & 1815.00  & 54 \tabularnewline
\hline 
59.7  & 46.0  & 2746.20  & 91 \tabularnewline
\hline 
60.0  & 37.7  & 2262.00  & 63 \tabularnewline
\hline 
60.6 & 46.0  & 2787.60  & 106 \tabularnewline
\hline 
81.3  & 38.3  & 3113.79  & 76 \tabularnewline
\hline 
83.2 & 41.1 & 3419.52 & 74\tabularnewline
\hline 
\end{tabular}
\end{table}
\begin{figure}[h] 
    \centering
    \includegraphics[width=0.6\textwidth]{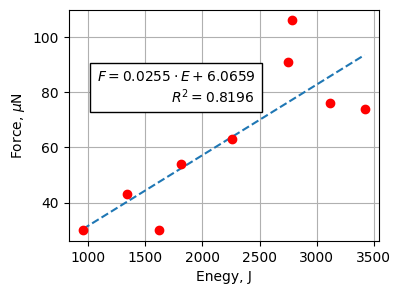} 
    \caption{RF resonator trust force vs energy experiment \cite{Ref31}}
    \label{fig:label} 
\end{figure}
\section{Mass increasing effect of inductance coil}
\label{sec:3}
If do not zero space curvature of electric field explains Biefeld{}-Brown effect  excellently, the same effect, but with increasing of device mass, must be observed in a magnetic field. Analogically to eq.~\eqref{eq:Eq21} electro-gravity mass equation for inductance coil looks like this
\begin{equation}
\label{eq:Eq22}
M_{eg}=-\frac{{{LI}}^{{2}}}{2} \alpha_g, 
\end{equation}
where ${L}$ is the inductance, which depends on area and amount of windings and coil length, and ${I}$ is the current. Equation eq.~\eqref{eq:Eq22} predict similar effects, observed in Biefeld-Brown experiments with a capacitor, but in opposite direction of changing of coil mass: for the fixed inductance ${L}$ increasing of current in coil must increase mass of the coil. 
\section{Conclusions}
\label{sec:4}
On the basis of the results obtained in this work the following
conclusions have been made:

1) Electro gravity mass equation is in good agreement with
Biefeld{}-Brown effect and fully explains all qualitative data observed
in this effect.

2) Electromagnetic gravity model predicts decreasing of gravity mass
effect in electric field and increasing of gravity mass in magnetic field.

3) Electromagnetic field gravity constant equals to ${\alpha_g = 2.60\pm 0.47E-9}$ [$\frac{s^2}{m^{2}}$]   and is the basic
constant of proposed electromagnetic gravity model.

4) The total curvature of electromagnetic wave is zero, but separate curvatures of electric and magnetic components of the wave haven't zero values.

\section{Acknowledgements}
I'd like to thank Gintaras Valiulis from Vilnius Univerity and Algis Dziugys from Lithuanian Energy Institute for discussions and notes, which let's definitely understand problem of new proposed electro-gravity model.

\appendix
\section*{Appendix: Discussion on classic derivation of electromagnetic energy-stress tensor}
According to \cite{Ref20} energy-stress tensor ${T^{ik}}$ of electromagnetic field could be derived from action principle
\begin{equation}
\delta S = \delta \int \Lambda\left(q , \frac{\partial q}{\partial x^i}\right) d V d t = 0.
\end{equation}
Variation of given action and determination it to zero leads to equation 
\begin{equation}
\label{eq:Ap1}
\frac{\partial T^{k}_{i}}{x^k} = 0,
\end{equation}
where
\begin{equation}
T^{k}_{i} = q_{,i} \frac{\partial \Lambda}{\partial q_{,k}} - \delta^{k}_{i} \Lambda.
\end{equation}
Equation ~\eqref{eq:Ap1} is equivalent to proposition, that it is obtained conservation low of 4-momentum vector ${P^i}$ 
\begin{equation}
\label{eq:Ap2}
P^i = \frac{1}{c} \int T^{ik} d S_k,
\end{equation}
where integration is making in all hyperplane. Description of ${T^{ik}}$, grounded on ~\eqref{eq:Ap2} is ambiguous, because every tensor described as 
\begin{equation}
\label{eq:Ap4}
T^{ik} + \frac{\partial}{\partial x^i} \psi^{ikl}, \ \ \ \psi^{ikl} = - \psi^{ilk}
\end{equation}
meets conservation low ~\eqref{eq:Ap1}. Trace of given energy-stress tensor with additional term 
\begin{equation}
\label{eq:Ap5}
T + \frac{\partial}{\partial x^i} \psi^{iil}
\end{equation}
generally isn't zero and must be calibrated with experiment.

The other rotational momentum conservation low specify energy-stress tensor to be symmetric
\begin{equation}
\label{eq:Ap3.0}
T^{ik} = T^{ki}
\end{equation}
and it could be reached with choosing of ${\psi^{ikl}}$. 

Now all this could be used to get ${T^{ik}}$ for Lagrangian of electromagnetic field
\begin{equation}
\Lambda = - \frac{1}{4 \mu_0} F_{kl} F^{kl},
\end{equation}
which gives 
\begin{equation}
\label{eq:Ap3}
T^k_i = -\frac{1}{\mu_0} \frac{\partial A^l}{\partial x_i} F^{kl} + \frac{1}{4 \mu_0} \delta^k_i F_{lm} F^{lm}, 
\end{equation}
where ${F_{kl}}$ is described as 
\begin{equation}
F_{kl} = \frac{\partial A_l}{\partial x^k} - \frac{\partial A_k}{\partial x^l}.
\end{equation}
Tensor in eq.~\eqref{eq:Ap3} isn't symmetric. The trace calculation of energy-stress tensors eq.~\eqref{eq:Ap3} gives
\begin{equation}
\label{eq:Ap6}
T^i_i = -\frac{1}{\mu_0} \frac{\partial \left(A^l F^{il}\right)}{\partial x_i}  + \frac{1}{\mu_0} F_{lm} F^{lm},
\end{equation}
which is true in electromagnetic field without charge, because ${\partial F^{il} / \partial x_i = 0}$. Equation ~\eqref{eq:Ap6} satisfy ~\eqref{eq:Ap5} transformation and finally the last equation could be rewritten as
\begin{equation}
\label{eq:Ap8}
T^i_i = \frac{1}{\mu_0} F_{lm} F^{lm} \neq 0.
\end{equation}
Symmetrization of ${T_k^i}$ for indexies ${i \neq k}$ could be reached, if the term would be added 
\begin{equation}
\label{eq:Ap7}
\frac{1}{\mu_0} \frac{\partial A^i}{\partial x_l} F^k_l = \frac{1}{\mu_0} \frac{\partial \left( A^i F^k_l \right)}{\partial x_l}
\end{equation}
It could be made too, because it is Eq.~\eqref{eq:Ap4} transformation. The result of folloing transformations is new symmetric tensor 
\begin{align}
T^{ik}_{\left( i = k \right)} &= -\frac{1}{\mu_0} \frac{\partial A^l}{\partial x^i} F^{kl}+ \frac{1}{4 \mu_0} \frac{\partial \left(A^l F^{il}\right)}{\partial x^i} + \frac{1}{4} g^{ik} T \\ 
T^{ik}_{\left( i \neq k \right)} &=  \frac{1}{\mu_0} \left( - F^{il} F^k_l + \frac{1}{4} g^{ik} F_{lm} F^{lm} \right) \label{eq:Ap71b} 
\end{align}
Transformations ~\eqref{eq:Ap7} applayed to whole ${T_k^i}$ tensor change both not diagonal elements and diagonal elements of energy-stress tensor, which trace after transformation become zero
\begin{equation}
\label{eq:Ap9}
T^i_i = 0.
\end{equation}
The answer to this question, which equation eq.~\eqref{eq:Ap8} or eq.~\eqref{eq:Ap9} is true, could be given just by experiment.


\begin{thebibliography}{}
%
%
\bibitem{Ref1}
Brown T.T., A Method of and an Apparatus or
Machine for Producing Force or Motion, U.K. Patent No. 00.311, 15
Nov. (1928)
\bibitem{Ref2}
Brown, T. T., How I Control Gravitation, Science and Invention, Aug. (1929) (reprinted in Psychic Observer, Vol. 37, No. 1, pp. 66, 67).
\bibitem{Ref3}
Brown, T. T., Electrostatic Motor, U.S. Patent 1.974.483, 25 Sept. (1934)
\bibitem{Ref4}
Brown, T. T., Electrokinetic Apparatus, U.S. Patent 2.949.550, 16 Aug. (1960)
\bibitem{Ref5}
Brown, T. T., Electrokinetic Transducer, U.S. Patent 3.018.394, 23 Jan. (1962)
\bibitem{Ref6}
Brown, T. T., Electrokinetic Generator, U.S. Patent 3.022.430, 20 Feb.
(1962)
\bibitem{Ref7}
Christensen, E. A., and M{\o}ller, P. S., Ion{}-Neutral Propulsion in Atmospheric Media, AIAA Journal, Vol. 5, No. 10, 1768{--}1773, (1967)
\bibitem{Ref8}
M.Tajmar, Biefeld{}-Brown Effect: Misinterpretation of Corona Wind Phenomena, AIAA Journal, Vol. 42, Nr. 2, pp. 315{}-318, (2004)
\bibitem{Ref9}
Brown, T. T., Electrokinetic Apparatus, U.S. Patent 3.187.206, 1 June (1965)
\bibitem{Ref10}
Talley, R. L., 21st Century Propulsion Concept, U.S. Air Force Astronautics Lab., Final Rept. AFAL{}-TR{}-88{}-031, April, (1988)
\bibitem{Ref11}
Talley, R. L., Twenty First Century Propulsion Concept, U.S. Air Force Propulsion Directorate, Final Rept. PL{}-TR{}-91{}-3009, May, (1991)
\bibitem{Ref12}
Williams, P. E., The Possible Unifying Effect of the Dynamic Theory, Los Alamos Scientific Lab., Rept. LA{}-9623{}-MS, Los Alamos, NM, May, (1983)
\bibitem{Ref122}
Hilbert, D. Die Grundlagen der Physik, Konigl. Gesell. d. Wiss. Gottingen, Nachr. Math.-Phys. Kl. 395-407 (1915)
\bibitem{Ref0}
Carroll, Sean M., Spacetime and Geometry, Addison Wesley (2004)
\bibitem{Ref13}
Einstein A., Die Grundlage der allgemeinen Relativit\"atstheorie, 
Annalen der Physik {49},(1916) . Retrieved on 2006{}-09{}-03
\bibitem{Ref14}
Einstein, A. Relativity: The Special and General Theory. New York: Crown, (1961).
\bibitem{Ref14_1}
Hawkin S.W., F.R.S and Ellis G.F.R., The large scale structure of space-time. Cambrige University Press, (1994).
\bibitem{Ref19}
R.C. Tolman,   "Relativity, thermodynamics and cosmology" , Clarendon Press  (1934)
\bibitem{Ref23}
Buehler D., R. Exploratory Research on the Phenomenon of the Movement of High Voltage Capacitors, Journal of Space Mixing Vol. 2, 1-22, (2004)
\bibitem{Ref20}
Landau, L., D., Lifshitz E., M. The Classical Theory of Fields : Volume 2. Oxford: Butterworth-Heinemann, (1980)
\bibitem{Ref31}
White Harold,   March Paul,   Lawrence James,   Vera Jerry,   Sylvester Andre,   Brady David and  Bailey Paul. Measurement of Impulsive Thrust from a Closed Radio-Frequency Cavity in Vacuum.  Journal of Propulsion and Power. Published Online:17 Nov 2016,  https://doi.org/10.2514/1.B36120
\end{thebibliography}


\end{document}